\documentstyle[preprint,pra,aps]{revtex}

\begin{document}

\draft

\tightenlines

\title{Rabi oscillations and macroscopic quantum superposition states}

\author{Marco Frasca}
\address{Via Erasmo Gattamelata, 3,
         00176 Roma (Italy)}

\date{\today}

\maketitle

\abstract{
A two-level atom interacting with a single radiation mode is considered, without the rotating-wave
approximation, in the strong coupling regime. It is shown that, in agreement with the recent results
on Rabi oscillations in a Josephson junction 
(Y. Nakamura, Yu. A. Pashkin and J. S. Tsai, Phys. Rev. Lett. {\bf 87}, 246601 (2001)), 
the Rabi frequency is indeed proportional to first kind
integer order Bessel functions in the limit of a large number of photons 
and the dressed states are macroscopic quantum superposition states. To approach this
problem analytically use is made of the dual Dyson series and the rotating-wave approximation.
}


\narrowtext

A recent experimental finding on Josephson junctions \cite{naka} has shown as Rabi oscillations
happen in strong electromagnetic fields for the two-level model. One of the main results
of this experiment was the proportionality of the Rabi frequency to first kind integer order
Bessel functions with the order given by the photon number involved in the transition.

The study of a two-level model in a cosine time-dependent perturbation in the strong coupling
regime \cite{fra1} proved that in this case Rabi oscillations involve only odd order first kind
Bessel functions and partially explains the results of Ref.\cite{naka}. On the same ground we
approached the problem of a two-level atom interacting with a single radiation mode in the
strong coupling regime in Ref.\cite{fra2}. In this paper we want to extend the results of
Ref.\cite{fra2} by discussing the case of the experiment of Ref.\cite{naka} with a large
number of photons involved. We will show that theory and experiment indeed agree.

The Hamiltonian we start with has the simple form as also given in Ref.\cite{naka}
(neglecting the tunneling contribution as not essential)
\begin{equation}
    H = \omega a^\dagger a +\frac{\Delta}{2}\sigma_3+g\sigma_1(a^\dagger+a) \label{eq:H}
\end{equation} 
being $\omega$ the frequency of the radiation mode, $\Delta$ the separation between the two
levels of the atom, $g$ the coupling between the radiation field and the atom, 
$\sigma_1$ and $\sigma_3$ the Pauli matrices and $a$ and $a^\dagger$ the
annihilation and creation operators for the radiation field. This apparently simple model
is not exactly solvable unless the rotating-wave approximation is done, in this latter case the
solution is exactly known and the model is then named the Jaynes-Cummings model\cite{Sch}.
The model (\ref{eq:H}) is able to describe the results of Ref.\cite{naka} when the coupling $g$ is large
with respect to the level separation.

To reach our goal we apply the duality principle in perturbation theory as given in Ref.\cite{fra3,fra4}.
The idea is to consider, contrarily to small perturbation theory as given e.g. in Ref.\cite{Mes},
as unperturbed Hamiltonian the term
\begin{equation}
    H_0=\omega a^\dagger a + g\sigma_1(a^\dagger +a).
\end{equation}
The Schr\"odinger equation ($\hbar=1$)
\begin{equation}
    H_0U_F(t)=i\frac{\partial}{\partial t}U_F(t)
\end{equation}
has the solution \cite{fra2}
\begin{equation}
   U_F(t)=\sum_{n,\lambda}e^{-iE_nt}|[n;\alpha_\lambda]\rangle\langle[n;\alpha_\lambda]|
	|\lambda\rangle\langle\lambda| \label{eq:uf}
\end{equation}
being $E_n=n\omega-\frac{g^2}{\omega}$, 
$\alpha_\lambda=\frac{\lambda g^2}{\omega}$ and
\begin{equation}
    |[n;\alpha_\lambda]\rangle=e^{\frac{g}{\omega}\lambda(a-a^\dagger)}|n\rangle
\end{equation}
a displaced number state \cite{kn},
$n$ an integer starting from zero the eigenvalue of the operator 
$a^\dagger a$ and $\lambda=\pm 1$ the eigenvalues of $\sigma_1$. These
states represent the dressed state for the system but our analysis complies
with the one given in Ref.\cite{fra4} by a dual Dyson series otherwise,
no Rabi oscillations can be obtained theoretically.

At this point we are able to write down a dual Dyson series for the Hamiltonian
(\ref{eq:H}) as
\begin{equation}
    U(t)=U_F(t)T\exp\left[-i\int_0^tdt'H_F(t')\right] \label{eq:dds}
\end{equation}
having put
\begin{equation}
    H_F(t)=U_F^\dagger(t)\frac{\Delta}{2}\sigma_3 U_F(t). \label{eq:H0}
\end{equation}
It is important to note that in the dual Dyson series (\ref{eq:dds}), as also happens
in the small perturbation case, when there is a resonance between the two-level
atom and the radiation field, perturbative terms appear that are unbounded in the
limit $t\rightarrow\infty$ and the perturbation series is useless unless we are
able to resum such terms, named ``secularities'' as in celestial mechanics,
at all order. This can be done e.g. by renormalization group methods \cite{fra5}
but here we limit the complexity of the mathematical analysis by simply doing
the rotating-wave approximation and ignoring any correction to it.

Let us look at the Hamiltonian (\ref{eq:H0}). It is easily realized that can
be rewritten in the form \cite{fra2}
\begin{equation}
    H_F=H_0'+H_1
\end{equation}
being
\begin{equation}
    H'_0=\frac{\Delta}{2}\sum_n e^{-\frac{2g^2}{\omega^2}}
	L_n\left(\frac{4g^2}{\omega^2}\right)
	\left[
	|[n;\alpha_1]\rangle\langle[n;\alpha_{-1}]||1\rangle\langle -1|+
	|[n;\alpha_{-1}]\rangle\langle[n;\alpha_1]||-1\rangle\langle 1|
	\right]
\end{equation}
being $L_n$ the n-th Laguerre polynomial \cite{grad} and
\begin{eqnarray}
    H_1=\frac{\Delta}{2}\sum_{m,n,m\neq n}e^{-i(n-m)\omega t}
	\left[
	\langle n|e^{-\frac{2g}{\omega}(a-a^\dagger)}|m\rangle
	|[n;\alpha_1]\rangle\langle[m;\alpha_{-1}]||1\rangle\langle -1|+
	\right.\nonumber \\
	\left.
	\langle n|e^{\frac{2g}{\omega}(a-a^\dagger)}|m\rangle
	|[n;\alpha_{-1}]\rangle\langle[m;\alpha_1]||-1\rangle\langle 1|
	\right].
\end{eqnarray}
The Hamiltonian $H_0'$ can be immediately diagonalized by the
eigenstates
\begin{equation}
    |\psi_n;\sigma\rangle=\frac{1}{\sqrt{2}}
	\left[
	\sigma|[n;\alpha_1]\rangle|1\rangle+
	|[n;\alpha_{-1}]\rangle|-1\rangle
	\right]
\end{equation}
with eigenvalues
\begin{equation}
    E_{n,\sigma}=\sigma\frac{\Delta}{2}e^{-\frac{2g^2}{\omega^2}}
	L_n\left(\frac{4g^2}{\omega^2}\right)
\end{equation}
being $\sigma=\pm 1$. Then we can see that each level of the atom develop
a band with an infinite subset of levels numbered by the integer number $n$.
The eigenstates can be seen as macroscopic quantum superposition states 
(sometimes named in the literature as Schr\"odinger cat states)\cite{Sch}. We will
prove that the two-level atom shows Rabi oscillations between these states. To prove
this result we look for a solution of the Schr\"odinger equation with the
Hamiltonian $H_F$ by taking
\begin{equation}
    |\psi_F(t)\rangle=\sum_{\sigma,n}e^{-iE_{n,\sigma}t}a_{n,\sigma}(t)|\psi_n;\sigma\rangle
\end{equation}
that gives the equations for the amplitudes \cite{fra2}
\begin{equation}
    i\dot{a}_{m,\sigma'}(t)=\frac{\Delta}{2}\sum_{n \neq m,\sigma}a_{n,\sigma}(t)
	e^{-i(E_{n,\sigma}-E_{m,\sigma'})t}e^{-i(m-n)\omega t}
	\left[\langle m|e^{-\frac{2g}{\omega}(a-a^\dagger)}|n\rangle\frac{\sigma'}{2}+
	\langle m|e^{\frac{2g}{\omega}(a-a^\dagger)}|n\rangle\frac{\sigma}{2}
	\right]. \label{eq:amp}
\end{equation}
At this stage we can apply the rotating-wave approximation. The resonance
condition is given by
\begin{equation}
	E_{n,\sigma}-E_{m,\sigma'}-(n-m)\omega=0
\end{equation}
and two Rabi frequencies are obtained. 
For interband transitions ($\sigma\neq\sigma'$) one has
\begin{equation}
	{\cal R}=\Delta|\langle n|\sinh\left[\frac{2g}{\omega}(a-a^\dagger)\right]|m \rangle|
\end{equation}
while for intraband transitions one has
\begin{equation}
	{\cal R'}=\Delta|\langle n|\cosh\left[\frac{2g}{\omega}(a-a^\dagger)\right]|m \rangle|.
\end{equation}
By using the relation
\begin{equation}
    \langle m|\exp\left[\frac{2g}{\omega}(a^\dagger-a)\right]|n\rangle=
	\sqrt{\frac{n!}{m!}}\left(\frac{2g}{\omega}\right)^{m-n}e^{-\frac{2g^2}{\omega^2}}
	L_n^{(m-n)}\left(\frac{4g^2}{\omega^2}\right)
\end{equation}
with  $L_n^{(m-n)}(x)$ an associated Laguerre polynomial \cite{grad}, it easy to
show that
\begin{equation}
    {\cal R}=\frac{\Delta}{2}\sqrt{\frac{n!}{m!}}\left(\frac{2g}{\omega}\right)^{m-n}e^{-\frac{2g^2}{\omega^2}}
	|L_n^{(m-n)}\left(\frac{4g^2}{\omega^2}\right)|[1-(-1)^{m-n}]
\end{equation}
and
\begin{equation}
    {\cal R'}=\frac{\Delta}{2}\sqrt{\frac{n!}{m!}}\left(\frac{2g}{\omega}\right)^{m-n}e^{-\frac{2g^2}{\omega^2}}
	|L_n^{(m-n)}\left(\frac{4g^2}{\omega^2}\right)|[1+(-1)^{m-n}]
\end{equation}
and then, for interband transitions one can have Rabi oscillations only between states differing by an odd number
and we write $m-n=2N+1$, while interband Rabi oscillations can happen only for states differing by an even number
and we write in this case $m-n=2N$. So, finally
\begin{equation}
    {\cal R}=\Delta\sqrt{\frac{n!}{(n+2N+1)!}}\left(\frac{2g}{\omega}\right)^{2N+1}e^{-\frac{2g^2}{\omega^2}}
	|L_n^{(2N+1)}\left(\frac{4g^2}{\omega^2}\right)|
\end{equation}
and
\begin{equation}
    {\cal R'}=\Delta\sqrt{\frac{n!}{(n+2N)!}}\left(\frac{2g}{\omega}\right)^{2N}e^{-\frac{2g^2}{\omega^2}}
	|L_n^{(2N)}\left(\frac{4g^2}{\omega^2}\right)|.
\end{equation}
We can interpret this Rabi oscillations as involving an effective number of photons $2N+1$ and $2N$ respectively
in the transitions.So, we can take $n$ to be very large and $N$ small or zero and this is in agreement with
the experiment described in Ref.\cite{naka}. This in turn means, in agreement with the experimental results,
\begin{equation}
    {\cal R}\approx\Delta\left|J_{2N+1}\left(\frac{4\sqrt{n}g}{\omega}\right)\right|
\end{equation}
and
\begin{equation}
    {\cal R'}\approx\Delta\left|J_{2N}\left(\frac{4\sqrt{n}g}{\omega}\right)\right|.
\end{equation}
where use has been made of the Stirling approximation for the factorial $n!\approx e^{-n}n^n\sqrt{2\pi n}$
at large $n$ and the equation \cite{grad}
\begin{equation}
    J_\alpha(2\sqrt{n}x)=e^{-\frac{x}{2}}\left(\frac{x}{n}\right)^\frac{\alpha}{2}L_n^\alpha(x)
\end{equation}
that holds in the limit of $n$ going to infinity, both for ${\cal R}$ and ${\cal R'}$.

To complete this paper, we want to show how Rabi oscillations emerge from eqs.(\ref{eq:amp}) in the limit
of a large number of photons involved, when we start taking as initial state e.g. $|0\rangle|g\rangle$
being $a|0\rangle=0$ and $\sigma_3|g\rangle=-|g\rangle$. Indeed, one has
\begin{equation}
    |0\rangle|g\rangle = \sum_{n,\sigma}a_{n,\sigma}(0)|\psi_n;\sigma\rangle
\end{equation}
being
\begin{equation}
    a_{n,\sigma}(0)=e^{-\frac{g^2}{2 \omega^2}}\left(\frac{g}{\omega}\right)^n
	\frac{1}{2\sqrt{n!}}[\sigma + (-1)^n]. \label{eq:amp0}
\end{equation}
For interband resonance, with $n-m=2N+1$, we get for $E_{n,(\sigma=1)}-E_{m,(\sigma'=-1)}=(2N+1)\omega$
with even $n$ and odd $m$ and $n>m$,
\begin{eqnarray}
    a_{m,-1}(t) &=& a_{m,-1}(0)\cos(\frac{\cal R}{2}t)-ia_{n,1}(0)\sin(\frac{\cal R}{2}t) \\
	a_{n,1}(t)  &=& a_{n,1}(0)\cos(\frac{\cal R}{2}t)-ia_{m,-1}(0)\sin(\frac{\cal R}{2}t)
\end{eqnarray}
that can be put in explicit form by the coefficients (\ref{eq:amp0}) giving
\begin{eqnarray}
    a_{m,-1}(t) &=& -e^{-\frac{g^2}{2\omega^2}}\left(\frac{g}{\omega}\right)^m
	\frac{1}{\sqrt{m!}}\cos(\frac{\cal R}{2}t)
	-ie^{-\frac{g^2}{2\omega^2}}\left(\frac{g}{\omega}\right)^n
	\frac{1}{\sqrt{n!}}\sin(\frac{\cal R}{2}t) \\
	a_{n,1}(t)  &=& e^{-\frac{g^2}{2\omega^2}}\left(\frac{g}{\omega}\right)^n
	\frac{1}{\sqrt{n!}}\cos(\frac{\cal R}{2}t)+ie^{-\frac{g^2}{2\omega^2}}\left(\frac{g}{\omega}\right)^m
	\frac{1}{\sqrt{m!}}\sin(\frac{\cal R}{2}t)
\end{eqnarray}
and finally
\begin{eqnarray}
    a_{m,-1}(t) &=& -e^{-\frac{g^2}{2\omega^2}}\left(\frac{g}{\omega}\right)^m
	\frac{1}{\sqrt{m!}}\left[\cos(\frac{\cal R}{2}t)
	+i\left(\frac{g}{\omega}\right)^{2N+1}\sqrt{\frac{m!}{(m+2N+1)!}}\sin(\frac{\cal R}{2}t)\right] \\
	a_{m+2N+1,1}(t)  &=& e^{-\frac{g^2}{2\omega^2}}\left(\frac{g}{\omega}\right)^m
	\frac{1}{\sqrt{m!}}\left[\left(\frac{g}{\omega}\right)^{2N+1}
	\sqrt{\frac{m!}{(m+2N+1)!}}\cos(\frac{\cal R}{2}t)+i\sin(\frac{\cal R}{2}t)\right].
\end{eqnarray}
In the limit of a very large number $m$ of photons we easily realize that
the Rabi frequency is
\begin{equation}
    {\cal R}\approx\Delta |J_{2N+1}\left(\frac{4\sqrt{m}g}{\omega}\right)|
\end{equation}
and we have the oscillating amplitudes
\begin{eqnarray}
    a_{m,-1}(t) &=& -e^{-\frac{g^2}{2\omega^2}}\left(\frac{g}{\omega}\right)^m
	\frac{1}{\sqrt{m!}}\cos(\frac{\cal R}{2}t) \\
	a_{m+2N+1,1}(t)  &=& e^{-\frac{g^2}{2\omega^2}}\left(\frac{g}{\omega}\right)^m
	\frac{1}{\sqrt{m!}}i\sin(\frac{\cal R}{2}t)
\end{eqnarray}
in agreement with the experimental results of Ref.\cite{naka}. It is interesting to
note that the probability to find the atom in one of the two levels is in any case
proportional to a Poisson distribution. Similar expressions can be obtained for
resonant intraband transitions ($\sigma=\sigma'$) and so we can have Rabi oscillations
with Rabi frequencies being proportional to odd or even order first kind Bessel
functions. The situation is quite different if instead of a second quantized radiation
field we use a classical cosine field \cite{fra5}: In this case we can have
Rabi frequency just with odd first kind Bessel functions. This is due to the 
disappearance of the band structure for a ``classical'' field 
and then, to the disappearance of the intraband transitions.

The above computation gives a strong theoretical support to the experimental findings
of Ref.\cite{naka}. On a different ground, we can state that the description through
dressed states as generalized in Ref.\cite{fra4} is sound. The experiment realized
by Nakamura et al., besides to be a first realization of a strongly perturbed two-level
system by a radiation a field, can be seen as the realization of oscillations between
macroscopic quantum superposition states. However, it is important to point out that decoherence is
observed whose source is to be identified in view of a practical use of Josephson
junction as gates for quantum computation. But, it is a fundamental result to have
proven experimentally the existence of a two-level system in a strong coupling regime.

In conclusion we have shown how the experimental findings in Ref.\cite{naka} can be
explained theoretically by the dual Dyson series and the generalized understanding
of dressed states described in Ref.\cite{fra4}.

I am in debt with Yasunobu Nakamura for giving me a preprint of their paper before the
publication.
	
\label{end}

\end{document}